\documentstyle[preprint,prl,aps,epsfig,latexsym]{revtex}
\def\slash#1{{\rlap{\hspace{.08em}/}#1}}

\input{epsf}

\begin{document}
\draft

\preprint{\vbox{\hbox{CALT-68-2080}}}

\title{Quantum solitons at strong coupling}

\author{I.~W. Stewart $^{1,2}$ 
and P.~G. Blunden $^{1,3}$}

\address{
$^1$ Department of Physics, University of Manitoba,
Winnipeg, Manitoba, Canada R3T 2N2 \\
$^2$ Department of Physics, California Institute of Technology,
Pasadena, CA 91125\\
$^3$ Institute for Nuclear Theory, University of Washington,
Seattle, WA 98195}

\maketitle

\begin{abstract}

We examine the effect of one loop quantum corrections on the formation of
nontopological solitons in a strongly coupled scalar-fermionic Yukawa
theory.  The exact one fermion loop contribution is incorporated by using a
nonlocal method to correct the local derivative expansion approximation (DE)
of the effective action.   As the Yukawa coupling is increased we find that
the nonlocal corrections play an increasingly important role. The
corrections cause the scalar field to increase in depth while maintaining
its size.  This increases the energy of the bag configuration, but this is
compensated for by more tightly bound fermionic states with lower energy. 
In contrast to the semi-classical picture without quantum corrections, the
binding energy is small, and the total energy scales directly with the Yukawa
coupling.  This confirms the qualitative behavior found in earlier work
using the second order DE, although the quantitative solutions differ.
\end{abstract}


In theories that support static nontopological soliton solutions at the
semi-classical level, it is of interest to ask what effect quantum
corrections may have on the form of the solution. In determining these
solutions, one must face the issue of self-consistency --- solving a set of
coupled nonlinear differential equations to find the field configuration
that minimizes the energy functional.  This makes it useful to have a local
expansion of the effective action to model these vacuum effects.
One such local method is to expand the one loop
effective action in momentum space about zero momentum.  For nontrivial
background fields this is an expansion in increasing orders of background
field derivatives, the so called derivative expansion (DE).  Such an
expansion has been considered in theories of finite nuclei
\cite{nuclei_ref}, QED \cite{qed_ref}, and Friedberg--Lee type soliton
models \cite{Bagger_Naculich,Perry:87,Wasson:91}.

Of considerable interest is whether the DE converges rapidly enough to
make it a useful tool.  The nature of the convergence has been studied by
several authors \cite{Li_P_W:87,Li_Perry,Blunden:89,Wasson_Koonin:91}.  In a
recent study \cite{Stewart_Blunden:96}, we have devised a method in $3+1$
dimensions that enables the DE contribution to the field equations to be
corrected for cases where it does not converge.  This allows the one loop
fermionic vacuum scalar density to be included exactly.  The method relies
on noticing that a DE of the fermion Green function $S(\bbox{x},
\bbox{x};\omega)$ works well for large loop energies $\omega$ and for
higher partial waves.  This method involves making a partial wave expansion
of the scalar vacuum density, and correcting smaller partial wave DE
contributions by an exact evaluation of the fermionic Green function.  
Details of the method and a discussion of its reliability can be found in
\cite{Stewart_Blunden:96}. 

In this article we consider the effect that vacuum corrections can have on
the self-consistent solutions of a soliton model. For simplicity, we
consider the nontopological soliton model of Bagger and Naculich
\cite{Bagger_Naculich}.  These authors solve for bound states of $N$
fermions with a Yukawa coupling to a dynamic scalar field, while including
the effects of the one loop fermion vacuum using the DE to second order in
derivatives of the scalar field.  However, it is not clear a-priori that the
DE is applicable for this model.  This is because the DE is an asymptotic
expansion in inverse powers of $m R$, where $m$ is the fermion mass and $R$
is typically the soliton size or surface thickness.  Thus, it is only
appropriate for sufficiently large
solitons with $R\gg 1/m$.  Hence the applicability of the DE will depend
on whether $1/m$, or some other scale in the Lagrangian, is instrumental
in determining the size of the soliton self-consistent solution.

The choice of Lagrangian density is \cite{Bagger_Naculich}
\begin{equation}
{\mathcal{L}} = \sum_{i=1}^N \bar\psi_i \left[ i\slash\partial - {g \over
\sqrt{N}} \phi \right] \psi_i + \frac{1}{2} ( \partial_\mu \phi)^2 -
{\mu^2\over 8 N v^2} (\phi^2 - N v^2)^2.\label{lagrangian}
\end{equation}
There are $N$ flavors of fermions in this model, and we work in the limit of
large $N$. The utility of the large $N$ parameterization is to validate the
semi-classical treatment of the scalar field.  In particular, the scalar
loop contributions are suppressed by a factor $1/N$.  The scalar field
has a nonzero vacuum expectation value (VEV), $\phi_\upsilon = \langle \phi
\rangle = \sqrt{N} \upsilon $, and the fermion mass is identified as $m =
g\upsilon$. Note that due to the presence of a Landau pole
\cite{Bagger_Naculich}, only models with $g \lesssim 30$ are physically
acceptable.  

To examine the model, we follow Bagger and Naculich \cite{Bagger_Naculich}
and consider the simple case of solitons with $N$ fermions that all appear
in the lowest single particle energy state.
The fermionic wave function then has the form
\begin{equation}
  \psi = {1\over r} \left( \begin{array}{c}\displaystyle{i G(r)}\\
     \displaystyle{ F(r) \bbox{\sigma}\cdot\hat{r}}
  \end{array} \right){\chi\over (4\pi)^{1/2}} , \label{psi}
\end{equation}
where $\chi$ is a Pauli spinor.  By rescaling the scalar field to
$\varphi = g \phi/\sqrt{N}$, the $N$ dependence in the Lagrangian density
will contribute simply as an overall factor $N$.  The effective action to
one loop order in fermionic fluctuations is $\Gamma_{\rm eff}[\varphi] =
\int d^4x\, {\mathcal{L}} +\Gamma_{\rm vac}[\varphi]$, where $\Gamma_{\rm
vac}[\varphi]= -i{\rm Tr}\ln( S^{-1}(\varphi)S(\varphi_\upsilon))$, and
$S(\varphi)=1/(i\slash\partial -\varphi)$ is the fermionic green function
operator in the background $\varphi$.  This gives
\begin{equation}
\Gamma_{\rm vac}[\varphi] = -i \int d^4x\ {\rm tr} \langle x |
  \ln (i\slash\partial -\varphi) - \ln (i\slash\partial -
  m) | x \rangle. \label{gammavac}
\end{equation}
The ground state configuration for static fields is
then the one that minimizes the energy functional
$E_{\rm tot}[\varphi] = - \Gamma_{\rm eff}[\varphi] / \int dt$.

After some standard manipulations of Eq.\ (\ref{gammavac}), the vacuum
energy can be written as \cite{Wasson:91,Li_Perry}
\begin{equation}
E_{\rm vac}[\varphi] = -{i \over 2 \pi} \int d^3 x\, \int d\omega\,\omega\,
{\rm tr} [\gamma_0 S(\bbox{x},\bbox{x};i\omega)],\label{evac}
\end{equation}
where
\begin{equation}
S(\bbox{x},\bbox{x};i\omega) = \langle \bbox{x}|
{1\over i \gamma_0 \omega + i \bbox{\gamma}\cdot\nabla - \varphi(x)}
|\bbox{x}\rangle \label{green}
\end{equation}
is the fermion Green function in the coordinate space representation.
The scalar vacuum density is obtained from the energy by taking the
functional derivative $\delta E_{\rm vac}/\delta\varphi$, giving
\begin{equation}
\rho_{\rm vac}(\bbox{x}) = -{1\over 2\pi} \int d\omega\,
  {\rm tr} [S(\bbox{x},\bbox{x};i\omega)].\label{rhovac}
\end{equation}
One can derive
the DE expressions for $E_{\rm vac}$ and $\rho_{\rm vac}$
by rewriting the Green function as
\begin{eqnarray}
S(\bbox{x},\bbox{x};i\omega) &=& \int {d^3 p\over (2\pi)^3}\,
\langle\bbox{x}|\bbox{p}\rangle\,
{1\over i \gamma_0 \omega - \bbox{\gamma}\cdot\bbox{p} -
\varphi(i\nabla_{\bbox{p}})} \,\langle\bbox{p}|\bbox{x}\rangle \nonumber\\
&=& \int {d^3 p\over (2\pi)^3}\,
{1\over i \gamma_0 \omega - \bbox{\gamma}\cdot\bbox{p} -
\varphi(\bbox{x}+i\nabla_{\bbox{p}})}, \label{greende}
\end{eqnarray}
and expanding about $\varphi(x)$. Of course, the above expressions for
$E_{\rm vac}$ and $\rho_{\rm vac}$ need to be regularized and renormalized
to make them finite. Dimensional regularization can be used with the 
appropriate renormalization conditions given in \cite{Bagger_Naculich}.

It is useful to rescale all fields and variables in terms of the fermionic
mass scale $m$, which we can set equal to 1. The effective energy of
the system can then be written as an energy per fermion in units of $m=g v$:
\begin{eqnarray}
  E_{\rm tot}[\varphi] &=& E_{\rm scalar}[\varphi] + \epsilon +
     E_{\rm vac}[\varphi],\\
   E_{\rm scalar}[\varphi] &=& {1 \over g^2} \int d^3 x\, \biggl[ \frac{1}{2}\,
   (\nabla \varphi)^2 + {\mu^2 \over 8} (\varphi^2-1)^2 \biggr], \\
   E_{\rm vac}[\varphi] &=&  {1\over 16 \pi^2} \int d^3 x\, \biggl[
 \frac{1}{2}\, (\varphi^2-1)\ (3 \varphi^2-1) - 
\varphi^4 \ln(\varphi^2) 
-  \ln(\varphi^2) (\nabla \varphi)^2 \biggr] + {\mathcal{O}}(\partial^4).
\end{eqnarray}
The DE expression for the vacuum energy has been expanded to second order
in derivatives of the scalar field \cite{Bagger_Naculich,Chan:85}.  The
fermion energy $\epsilon$ is found by solving the eigenvalue equation $
(-i \bbox{\gamma}\cdot\nabla + \varphi) \psi = \epsilon \gamma_0 \psi$. 
Minimizing the energy functional, $E_{\rm tot}[\varphi]$, and using Eq. 
(\ref{psi}), allows us to write the field equations as
\begin{eqnarray}
  {d G \over dr} &=& {G \over r} + (\epsilon + \varphi) F,\label{field1}\\ 
  {d F \over dr} &=& -{ F \over r} - (\epsilon - \varphi) G, \label{field2} \\
   \nabla^2 \varphi + { \mu^2 \over 2 } ( \varphi - \varphi^3 ) &=&
       g^2 ( \rho_{\rm val}\ +\ \rho_{\rm vac} ),  \label{field3}
\end{eqnarray}
subject to the normalization constraint
\begin{equation}
\int_0^\infty dr (G^2 + F^2) = 1.\label{norm}
\end{equation}
The source terms are
\begin{eqnarray}
 \rho_{\rm val} &=&  {1 \over 4 \pi r^2 } (G^2 - F^2), \label{source1} \\
   \rho_{\rm vac} &=& \rho_{\rm vac}^{\rm LDA}\ + \rho_{\rm vac}^{\rm DE,2}\ +
 {\mathcal{O}}(\partial^4), \label{source2}\\
 \rho_{\rm vac}^{\rm LDA} &=&  -  \ {1 \over 4 \pi^2 }\ \left[ \varphi^3
\ln(\varphi^2) - \varphi^3 + \varphi \right], \label{source3} \\
\rho_{\rm vac}^{\rm DE,2} &=& {1 \over 8 \pi^2}\ \left[ {1 \over \varphi}
\left(\nabla\varphi \right)^2 + \ln(\varphi^2) (\nabla^2 \varphi) \right].
\label{source4}
\end{eqnarray}

As written, these equations facilitate trying different levels of
approximation to the fermionic vacuum in the self-consistent solution. 
Setting $\rho_{\rm vac}=0$ corresponds to the semi-classical approximation
(CL), where vacuum corrections are ignored.
Keeping the first term in (\ref{source2}) will be referred to as the
local density approximation (LDA), as this term gives the exact one loop
result for a spatially uniform scalar field.  Including the second term in
(\ref{source2}) will be referred to as derivative expansion (DE)
approximation.  The density $\rho_{\rm vac}$ has only been expanded to
second order --- the same order as the differential equation (\ref{field3})
for $\varphi$.  In our experience, attempting to include terms with higher
order derivatives in self-consistent calculations tends to make the
solutions numerically unstable.  We discuss below a method for evaluating
$\rho_{\rm vac}$ exactly.

It was noted in \cite{Bagger_Naculich} that for $g\rightarrow\infty$,
the soliton solution satisfies a constraint that becomes independent of
$g$.  To see this we note that all source terms enter with the
same power of $g$ in (\ref{source1}-\ref{source4}), and Eqs.\
(\ref{field1}) and (\ref{field2}) are $g$-independent, so with the above 
notation this constraint takes the simple form
\begin{equation}
 \rho_{\rm val}(r)\ +\ \rho_{\rm vac}(r) = {\mathcal{O}}(1/g^2).
\label{constraint}
\end{equation}
Thus a large amount of cancellation is required between these source terms
for large $g$.  This restricts the form of the solutions, but does not
necessarily enforce an asymptotic $g$-independent shape, as claimed in
\cite{Bagger_Naculich}.  The reason is that a family of different functions
may exist that satisfy Eq.\ (\ref{constraint}), so that the true solution
is determined by the ${\mathcal{O}}(1/g^2)$ parts.  

The field equations (\ref{field1}-\ref{field3}) agree with those of Bagger
and Naculich \cite{Bagger_Naculich}.  Although these authors checked the
convergence of the DE by examining the relative size of the fourth order
terms in the expansion, they did so only for the {\em energy} functional
using a fixed scalar profile.  The convergence, however, may be quite 
different for terms in the dynamical
equations of motion.  To see this, consider a fixed background 
field of the form
\begin{equation}
 \varphi(r) = 1 - { a (1+f) \over e^{b\,r^2} + f e^{-b\, r^2}}\ .
\label{phiback} 
\end{equation}
At values of $a=0.5$, $b=0.16$, and $f=0.8$ we obtain a scalar field that
roughly corresponds to the quantum soliton solution at $g=25$ found by
Bagger and Naculich \cite{Bagger_Naculich}.  Table \ref{table_bn1} shows the
DE contributions to the energy and the density up to fourth order.
Expressions for these fourth order contributions can be found in the
Appendix. It can be seen that although the fourth order terms do not make a
substantial contribution to the energy, this is not the case for the density
(which is shown for $r=0$ in the table).  For the self-consistent solution
the convergence is worse as $g$ increases and the size of the scalar
bag solution shrinks.

Our purpose therefore, is to reconsider this model and make use of the
correction method described in \cite{Stewart_Blunden:96} to
account for the fermionic vacuum in an exact manner.  This involves
extending a scheme discussed by Wasson and Koonin
\cite{Wasson:91,Wasson_Koonin:91} for calculations in one spatial
dimension. In three spatial dimensions, an exact, or ``brute-force''
calculation involves making a partial wave expansion of the Green function
(\ref{green}),
\begin{equation}
S(\bbox{x},\bbox{x}';i \omega) = {1 \over r r'} \sum_{\kappa,m}
 S_\kappa (r,r';i \omega)
 \otimes {\mathcal{Y}}_{\kappa m}(\hat{x})
 {\mathcal{Y}}^\dagger_{\kappa m}(\hat{x}').
\label{greenpwe}
\end{equation}
Each partial wave Green function obeys the equation
\begin{equation}
  \left( \begin{array}{cccc}
    i \omega -\varphi(r)  & \displaystyle{-{d\ \over dr} + {\kappa\over r}}
 \\  \displaystyle{-{d\ \over dr} - {\kappa\over r}} & -i \omega -\varphi(r) 
  \end{array} \right) S_\kappa(r,r';i \omega) = \delta(r-r') \label{rdirac1},
\label{greeneqn}
\end{equation}
which can be solved numerically \cite{Li_W_Perry}.  A similar partial wave
expansion can be made of the DE approximation to the Green function, Eq.\
(\ref{greende}), giving $S_\kappa^{\rm DE}$ \cite{Li_Perry}.  The sum of all
partial wave DE contributions to the density will reproduce the expressions
(\ref{source3}-\ref{source4}).  

Wasson and Koonin \cite{Wasson:91,Wasson_Koonin:91} pointed out that the DE
approximation to the Green function works well for large loop energies
$\omega$ (and similarly also for large angular momenta $\kappa$ in three spatial
dimensions).  Therefore one can use the DE as a sophisticated extrapolation
procedure for accelerating the convergence of the brute-force method.  The
exact Green function contribution can be calculated up to
some $\kappa_{\rm max}$, and the DE can be used to calculate the contribution
for the remaining partial waves.  An equivalent method is to use the full DE
result, and correct the low energy and low partial wave terms using the
difference between the exact and DE expressions for $S_\kappa$.  This
improves the convergence of the energy integral in (\ref{rhovac}).  Hence we
put
\begin{eqnarray}
   \rho_{\rm vac} =  \rho_{\rm vac}^{\rm LDA}\ + \rho_{\rm vac}^{\rm DE,2}\ +
 \ \Delta\rho_{\rm corr}[\varphi],
\label{rho}
\end{eqnarray}
where the density correction is given by
\begin{eqnarray}
  \Delta\rho_{\rm corr}[\varphi] &=& \sum_{\kappa=1}^{\infty}
\lim_{\Lambda_\kappa\rightarrow \infty}\Delta\rho_\kappa(\Lambda_\kappa,r),
\label{D_corr_cut} \\
\Delta\rho_{\kappa}(\Lambda_\kappa,r) 
 &=& - {\kappa \over \pi^2}  \int_{0}^{\Lambda_\kappa} d\omega \biggl[
 {1 \over r^2} \Re \,{\rm tr} [S_\kappa(r,r; i \omega)]  + 
\Upsilon^1_\kappa\varphi - \frac{1}{2}  \Upsilon^2_\kappa (\nabla^2\varphi)
\qquad\quad \nonumber\\
& &\qquad - \frac{2}{3}  \Upsilon^3_\kappa \varphi^2 (\nabla^2\varphi) 
- \frac{5}{3} \Upsilon^3_\kappa \varphi (\nabla\varphi)^2
- 2 \Upsilon^4_\kappa \varphi^3 (\nabla\varphi)^2 \biggr]
\label{Cor_D},
\end{eqnarray}
where
\begin{eqnarray}
\Upsilon^n_\kappa(r) &=& {1 \over 4 \pi r^2} [ \Delta_{\kappa-1}^n(r) +
\Delta_\kappa^n(r) ] \label{DE_pw},\\
\Delta_\kappa^n(r) &=& - {1 \over (n-1)!} \left( {1 \over 2 z_\omega} {d
\over d z_\omega} \right)^{n-1} [z_\omega r^2 i_\kappa(z_\omega r)
k_\kappa(z_\omega r)],\label{DEpwee}
\end{eqnarray}
\mbox{$z_\omega = (\omega^2+ \varphi^2)^{1/2}$}, $\varphi=\varphi(r)$, and
$i_\kappa$, $k_\kappa$ are the modified spherical Bessel functions of
order $\kappa$.  The utility of this method is that one only needs to
include in $\Delta\rho_{\rm corr}$ as many partial waves as are needed to
achieve
the desired accuracy.  Also, $\Delta \rho_{\rm corr}$ is a finite quantity
that is independent of renormalization.  The renormalization counterterms
only appear in the DE expressions (\ref{source3}-\ref{source4}).  The
details of deriving this correction are described in
\cite{Stewart_Blunden:96}.

Numerically, it is useful to be able to treat our equations entirely as a
boundary value problem.  This can be done by treating
$\epsilon$ as a field, and also introducing a field $\chi$ for the
auxiliary equation (\ref{norm}), so that
\begin{eqnarray}
  {d\epsilon \over dr} &=& 0, \quad
  {d\chi \over dr} = G^2 + F^2. \label{epschi}
\end{eqnarray}
The boundary conditions for the entire system of equations are then
\begin{eqnarray}
   \left. {d\varphi(r) \over dr}\right|_{r \rightarrow 0} &=& 0, \hspace{1.2in}
\left. {d \varphi(r) \over dr}\right|_{r \rightarrow R} =
\left(\mu + \frac{1}{R}\right) [1 - \varphi(R)], \nonumber \\
  \left. { F(r) \over r G(r)}\right|_{r \rightarrow 0} &=& \frac{1}{3}
[\varphi(0) - \epsilon],  \hspace{.84in}
   {F(R) \over G(R)} = - \left[ {\varphi(R) - \epsilon \over \varphi(R) +
\epsilon} \right]^{1/2},  \hspace{.7in} \label{BNbc}  \\
   \chi(0) &=& 0,   \hspace{1.6in}  \chi(R) = 1,\nonumber
\end{eqnarray}
where $R$ is large compared to the length scale of the problem. To solve
our equations we make use of the program COLNEW by
Ascher and Bader \cite{Ascher_Bader}.
Unfortunately, when the correction (\ref{Cor_D}) is included we cannot
simply use the COLNEW routine as written, since our equations are actually
integro-differential equations.  To facilitate this, COLNEW was
modified so that the scalar field solution and its derivative can be
extracted at intermediate stages of the calculation to evaluate the
density correction.  The calculation is then
internally iterative, with $\Delta\rho_{\rm corr}$ treated as a source term
that is reevaluated as needed with each internal iteration of the COLNEW
code. 

We now examine the results when using different levels of approximation to
the vacuum.  We use the schemes discussed above in the following
order: CL, LDA, DE, and EX.  Here EX will refer to the exact solution, which
includes $\Delta\rho_{\rm corr}$ (\ref{D_corr_cut}).  Our goal is to quantify
the effect of this correction on the self-consistent solution for various
couplings.  The choice $g=10$, $\mu=1/10$ will serve as an example.  (As the
energy scale $m$ already includes an implicit dependence on $g$, setting
$\mu = 1/g$ simply fixes the scalar field mass at the VEV value.)

\begin{figure}
\centerline{\epsfysize=5.37in  \epsfbox{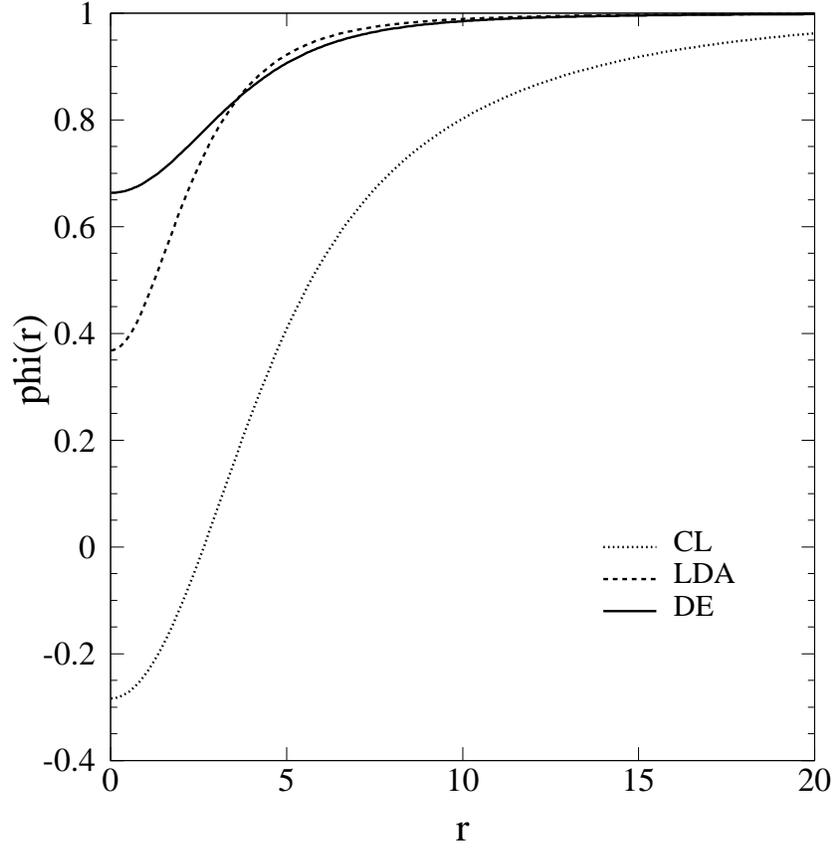}}
\caption{Self-consistent scalar field solutions of the Bagger-Naculich model
($g=10$, $\mu=1/10$) for semi-classical (CL),
local density (LDA), and derivative expansion (DE)
approximations to the vacuum densities.}
\label{fig_bn1}
\end{figure}

For the CL, LDA , and DE approximations the self-consistent solutions for
$\varphi$ are shown in Fig.\ \ref{fig_bn1}.  We see that adding the LDA term
reduces the depth and width of the scalar field, so that a zero no longer
appears.  The total energy of the solution, $E_{\rm tot}=.960 m$ (LDA), has
dramatically changed from that of the semi-classical approximation, $E_{\rm
tot}=.620 m$ (CL).  This behavior, which was identified by
\cite{Bagger_Naculich} at the DE level, occurs even before the derivative
terms are included.  However, for $g > 10.4$ the LDA approximation gives no
solution.  Looking at the expression for the LDA contribution in
(\ref{source3}), we see that this function is odd with respect to the scalar
field.  Increasing the coupling drives the scalar field deeper, so that near
$g=10.4$ the solution becomes negative at $r=0$, and the LDA contributes
with the opposite sign.  At this point the equations no longer support a
solution as it is impossible to satisfy the constraint (\ref{constraint}). 
However, when the DE terms are included the vacuum density is not forced to
change sign as $g$ is increased, and a soliton solution exists.  In Fig.\
\ref{fig_bn1} we see that the shape of the scalar field for $g=10$ is even
shallower for the DE than for the LDA.  An interesting feature to note is
that under the DE the total energy, $E_{\rm tot}=.967m$ (DE), is relatively
unchanged from the LDA result.  This occurs even though the Fermi level has
changed from $\epsilon=.863 m$ (LDA) to $\epsilon=.908 m$ (DE), and is made
possible due to a corresponding decrease in the energy contribution from the
scalar field, $E_{\rm scalar}+E_{\rm vac}$.  

\begin{figure}
\centerline{\epsfysize=5.37in  \epsfbox{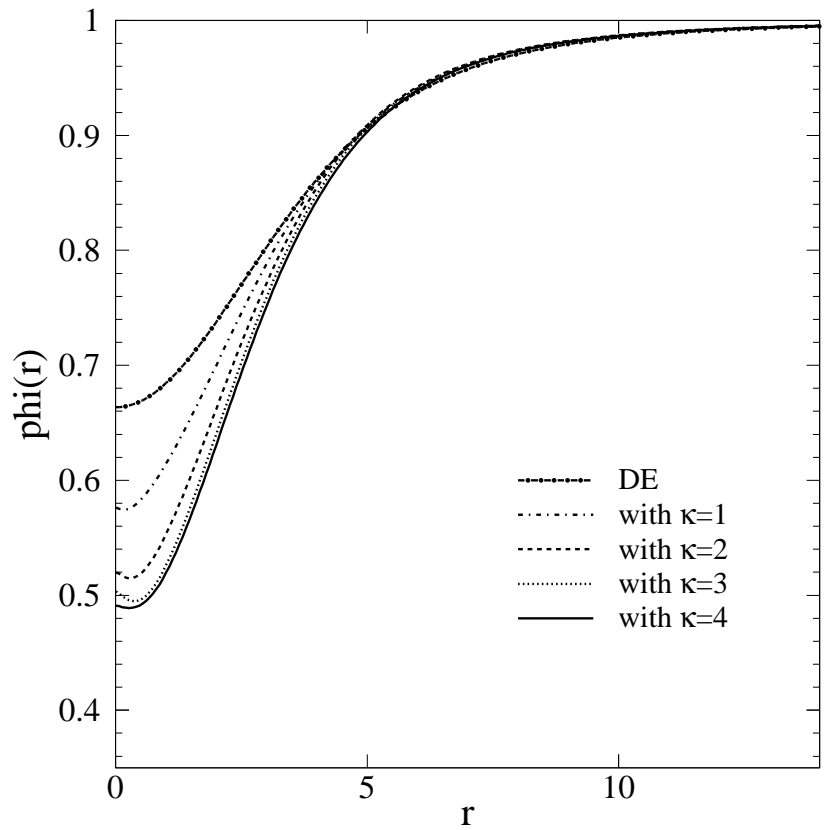}}
\caption{Self-consistent scalar field solutions with increasing number of
terms in the partial wave correction series. $g=10$, $\mu=1/10$.}
\label{fig_g10_pw}
\end{figure}

Now examine what happens when the density correction terms are included.  In
Fig.\ \ref{fig_g10_pw} the self-consistent scalar field solutions are shown
for calculations including an increasing number of partial wave corrections. 
The exact vacuum density favors a deeper scalar field than the DE.  The fifth
partial wave has not been included in the figure since the field in this case
is found to be indistinguishable from that with four partial waves.  Looking
at the sequence of scalar fields in Figs.\ \ref{fig_bn1} and \ref{fig_g10_pw}
we see a manifestation of the fact that terms in the derivative expansion
display oscillatory convergence \cite{Li_W_Perry}.   Note that after
including the $\kappa=1$ term in the correction (\ref{Cor_D}) the correct
vacuum density is specified at the origin as higher terms are zero there. 
However, the $\kappa > 1$ terms may still affect the self-consistent
solution.  In fact, we see in Fig.\ \ref{fig_g10_pw} that the scalar field
becomes deeper as subsequent terms in the correction are included.  The
corrected solution gives $\epsilon=.857 m$ (EX), and $E_{\rm tot}=.976 m$
(EX).  The energy of the scalar field is twice the DE result, and the Fermi
level has dropped away from $m$.  However, the total energy remains fairly
stable, rising only slightly above the DE result.  

\begin{figure}
\centerline{\epsfysize=5.37in  \epsfbox{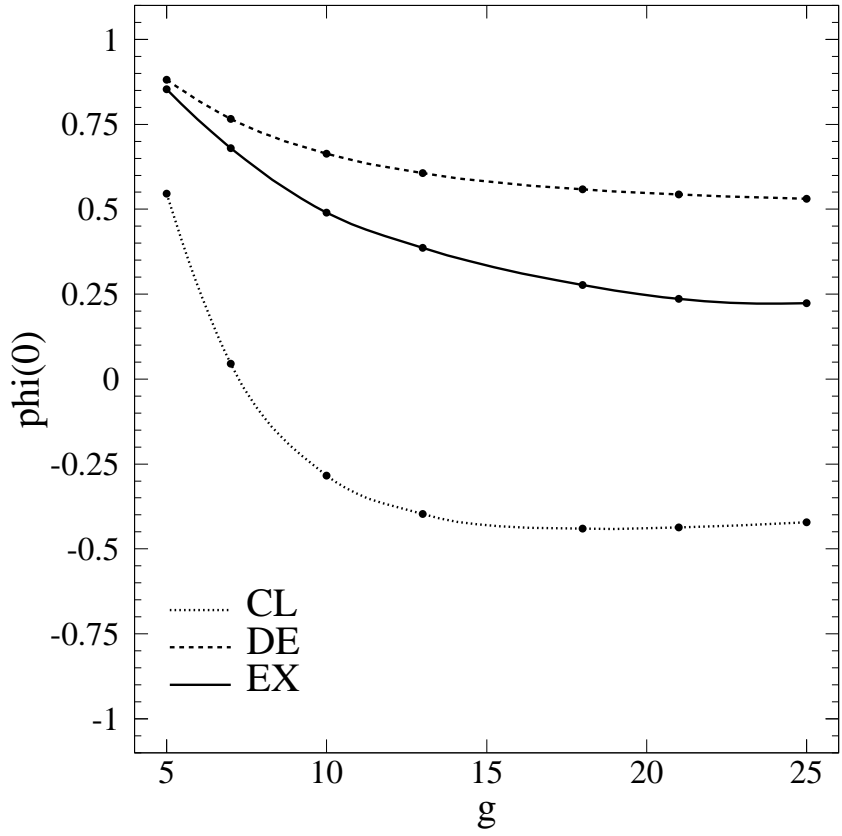}}
\caption{Depth of the self-consistent scalar field solutions $\varphi(r=0)$
for different values of the coupling $g$, and $\mu=1/g$.  Results are
shown for the semi-classical (CL), derivative expansion (DE), and exact
(EX) solutions.}
\label{fig_dvsg}
\end{figure}

\begin{figure}
\centerline{\epsfysize=5.37in  \epsfbox{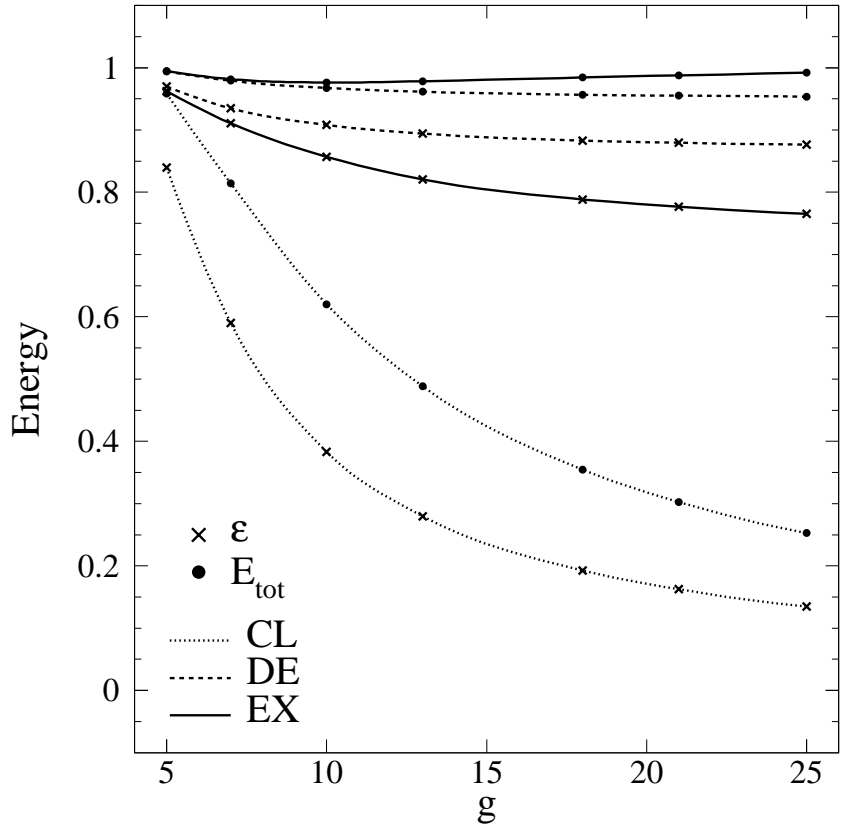}}
\caption{Self-consistent fermion ($\epsilon$) and total soliton ($E_{\rm tot}$)
energies for different values of the coupling $g$, with $\mu=1/g$.}
\label{fig_erung}
\end{figure}

To get the EX correction it is sufficient to use partial waves up to
$\kappa=4$, except at large $g$ where $\kappa=5$ is needed.  In Fig.\
\ref{fig_dvsg} we see that for other values of $g$ there is also a decrease
in the depth of the scalar field using the exact calculation.  The exact
results agree with the DE calculation for small coupling as expected.  As $g$
is increased the depth of the bag decreases.  For $g>20$ values of the
wave function $\varphi(r)$ near $r=0$ are numerically uncertain by about
$0.02$.  This uncertainty is a symptom of the constraint (\ref{constraint}),
but in no way affects the value of the energies.  In Fig.\ \ref{fig_erung} we
show how including the correction affects the energies at different $g$. 
>From the fermion energy, $\epsilon$, we see that the corrections affect the
solution for all $g>5$.  As the coupling increases we have found that not
only does the width of the bag shrink, but the depth increases.  Since this
leads to larger derivatives, the energy of the scalar field increases. 
Correspondingly, fermions may be more tightly bound in the bag so that the
Fermi energy level decreases.  When the fermionic and scalar field energies
are added to obtain the total energy, the sum gives an answer that is closer
to the DE result than either of the components separately.  This occurs at
all $g$.  In fact, it is surprising to find that the corrected total energy
remains fairly constant as the coupling is increased, exhibiting nearly the
same scaling as the DE total energy.  Recall that our energy is scaled in
terms of the fermionic mass $m=g \upsilon$, so that relative to $\upsilon$
the energy scales directly proportional to the coupling.  Thus the claim that
such scaling behavior is universal \cite{Naculich:92} is supported by our
calculations.

In summary, we have found a way of correcting the DE approximation for
fermionic vacuum fluctuations while still demanding a self-consistent
solution. There remain several ways in which the results of the
calculation done here may be made more general.  To include different
fermionic energy levels it would likely be more economical to use the
COLNEW routine to solve only the scalar equation, while solving the
eigenvalue problem for the relevant fermionic wave function components
using standard Runge-Kutta techniques.  If one takes for granted some
validity in the one loop approximation for finite $N$ and large coupling,
it should also be possible to include the scalar fluctuations at the
one loop level.  It is expected that these fluctuations would contribute
a source density term with the opposite sign to the fermionic
fluctuations \cite{Li_Perry}.  However, it is likely that a similar
density correction calculation would be necessary for the scalar loops to
ascertain that this contribution to the true scalar vacuum was also
included correctly.  

\acknowledgements
One of us (P.G.B.) would like to thank the Institute for Nuclear Theory
and the Nuclear Theory group at the University of Washington for their
hospitality and support during his sabbatical leave. I.W.S. would like to
thank Martin Gremm, Anton Kapustin, and Mark Wise for their comments.  This
work was also supported in part by the Natural Sciences and Engineering
Research Council of Canada, and by the U.\ S.\ Department of Energy under grant
DE-FG03-92-ER40701.

\appendix
\section{Fourth order Derivative Expansion of the Energy and Density}

The expressions for the energy and density at fourth order, which were
used in Table \ref{table_bn1}, are
\begin{eqnarray}
  E_{\rm vac}[\varphi] &=&  -  {1 \over 160 \pi^2} \int d^3 x\,  \biggl[
    { (\partial^2 \varphi )^2 \over \varphi^2} -
    {  11 (\partial_\alpha \varphi)^2 (\partial^2\varphi) \over 9 \varphi^3 } +
    {  11 (\partial_\alpha \varphi)^4 \over 18 \varphi^4 } \biggl], \\
  \rho_{\rm vac}^{DE,4} &=& - {1 \over 80 \pi^2 \varphi^2}
 \biggl[ \partial^4\varphi
 - {25  (\partial_\alpha\partial^2\varphi) (\partial^\alpha\varphi) \over
      9\varphi}
 - {11 (\partial^2\partial_\alpha\varphi) (\partial^\alpha\varphi) \over
      9\varphi} 
   - {16 (\partial^2\varphi)^2 \over 9 \varphi} \nonumber\\
& &
 - {11 ( \partial_\alpha\partial_\beta\varphi )^2 \over 9 \varphi}
 + {43 (\partial_\alpha\varphi)^2 (\partial^2\varphi) 
      \over 9 \varphi^2}  
  + {44 (\partial_\alpha\partial_\beta\varphi) (\partial^\alpha\varphi) 
    (\partial^\beta\varphi) \over 9 \varphi^2) }
 - {11 (\partial_\alpha\varphi)^4 \over 3 \varphi^3 }  \biggr]. \label{BNe4e}
\end{eqnarray}
Covariant notation is used for convenience.

\begin{table}[t]
\caption{Contributions from different orders of the DE series for the
energy and the density.  Results are shown for a fixed background field
$\varphi(r)$ with $a=0.5$, $b=0.16$, and $f=0.8$.
Units are given in terms of the scale $m$, as indicated. }
\rule{0in}{2ex}
\begin{center}
\begin{tabular}{lddd}  
& LDA & 2nd order & 4th order  \\ \hline
$E_{\rm vac}$  ($ 10^{-2} m$)   & 5.909 & 2.001  & $-$0.636 \\
$\rho_{\rm vac}(r=0)$ ($ 10^{-3} m^3$) & $-$5.109 & $-$0.936  & $-$3.732 \\  
\end{tabular}
\end{center}
\label{table_bn1}
\end{table}

\newpage

\newpage

\newpage

\newpage

\end{document}